# The Case for High-Resolution Infrared Spectroscopy with the Habitable Worlds Observatory

Daniel Jaffe[1], Gregory Mace[1], Erica Sawczynec[1], Ueejeong Jeong[1], Caroline Morley[1]
[1]University of Texas at Austin & McDonald Observatory, Austin, TX
dtj@austin.utexas.edu

## Abstract

A high-resolution near-IR spectroscopy capability on the Habitable Worlds Observatory (HWO) could strongly and efficiently advance many of the mission's goals. The technical barriers that made such a capability unfeasible on previous missions have largely been eliminated. Many HWO science case development documents require high spectral resolution in the IR and others would benefit significantly from it. High resolution improves the detectability of weak, unresolved features, aids identification of those features and provides additional information about radial velocity and line shape. It will be significantly easier to remove contaminating stellar features from high-resolution data. Silicon diffractive optics, immersion gratings and grisms, together with the new generation of low-noise, low dark-current avalanche photodiode arrays, make it possible to design a very compact high-resolution spectrograph that can cover the entire 1.1-2.0 μm band in a single exposure that would realize all of these advantages. We outline here the case for such an instrument and the technology development pathway needed to mature it in preparation for the HWO mission.

## Introduction

High-resolution near-IR spectroscopic capabilities can contribute to the attainment of Habitable Worlds Observatory's goals. Recent advances in disperser and detector technologies offer a path to making high resolution IR both capable and practical. HWO has the primary objective to identify, image, and characterize more than 25 potentially habitable worlds. By targeting the UV, visible and near-infrared from space, HWO will be able to observe other worlds without the obstruction of Earth's atmosphere.
A number of Science Case Development Documents (SCDDs) rely on high-resolution near-infrared spectroscopy: High-resolution UV-to-NIR Characterization of Exoplanet Atmospheres (Cubillos et al. 2026), Detecting and Characterizing the Magnetic Field of Exoplanets (Strugarek et al. 2026), Characterizing the Dynamics and Chemistry of Transiting Exoplanets (Wakeford et al. 2025) and White Dwarfs as Probes of Extrasolar Planet Compositions and Fundamental Astrophysics (Xu et al. 2025). Many other SCDDs identify near infrared spectroscopy as a fundamental requirement with breakthrough science occurring at medium-to-high resolution. NIRSpec on JWST provides near-infrared spectroscopy with resolving power R < 3600 (Jakobsen et al. 2022), which the SCDDs above identify as state-of-the-art, but insufficient for scientific progress with HWO.

Low resolution spectra present significant problems in analysis that high-resolution results can remove. It can be impossible to detect weak features from important species if stronger lines of more common species lie nearby. The identification of spectral features can be fraught with ambiguity. The lower information content in low resolution spectra makes it more difficult to remove residual spectral contamination by the host star.

In the past decade, spectral information about exoplanets has come from low to moderate resolution observations from space and from moderate to high spectral resolution observations from the ground. The ground-based observations have added significant information about the structure and dynamics of hot exoplanet atmospheres and have provided initial determinations of isotopic abundances (Snellen 2025, and references therein). Arguments against high spectral resolution observations from space have rested on lower detectability of the continuum, overcoming detector noise and dark current, lower throughput and instrument size and weight. We outline the science arguments for high resolution at 1.1 μm to up to 2.0 μm and show how technical developments in spectrograph components, spectrograph design, detectors and data analysis techniques make it possible to have the science advantages of high spectral resolution without the penalties in wavelength coverage and sensitivity that discouraged high-resolution on previous missions.

## A. Advantages of High-Resolution Spectroscopy

High-resolution spectra offer significant advantages in sensitivity and information content. Figure 1 compares high- and low-resolution observations. The central panel shows an observed transmission spectrum of the Earth's atmosphere from 1.5 to 1.7 μm at $R = \lambda/\Delta\lambda = 45,000$ (from the Immersion Grating Infrared Spectrometer, IGRINS, in black) and the same spectrum smoothed to R=2,000 (in yellow, note that high resolution for NIRSpec on JWST is R~2700 at this wavelength). Large numbers of molecular features (principally $H_2O$, $CH_4$, and $CO_2$) dominate the spectrum. The features are easy to identify at high resolution but it is only because many of these molecules produce groups of lines that are closely spaced in wavelength that they are also detectable and identifiable at low resolution.

If we ignore systematic effects and consider the light that passes through the spectrograph slit, we can write the signal to noise ratio for a single, unresolved line as:

$$SNR = \frac{\eta_s n_{line}}{\left(\eta_s(n_{chan}+n_{bg}-n_{line})+m(n_{dark}+n_{read}^2)\right)^{1/2}} \qquad (1)$$

Where $\eta_s$ is the throughput of the spectrograph, $n_{line}$ refers to the number of photons taken out by an absorption line, $n_{chan}$ is the number of source photons and $n_{bg}$ the number of background photons (from all sources combined) present in the continuum in a single resolution element, $m$ is the number of pixels per resolution element, $n_{dark}$ is the number of electrons from dark current per pixel and $n_{read}$ is the read noise per pixel.

**Raw detectability:** For the moment, we assume a throughput of 100% and consider only source photon noise. Let us assume that there are 100 detected photons per R = 45,000 resolution element (2250 photons per R = 2,000 element) in the continuum. If we consider an individual narrow feature with a depth of 35% at R=45,000 (like some of the features in Figure 1), the SNR will be $35/(65)^{1/2}$ =4.3 while at R=2,000, it would be $35/(2250)^{1/2}$ =0.7, about 6X lower. In the limiting case of weak features, the SNR advantage of high resolving power is

$$\left(\frac{\eta_H R_H}{\eta_L R_L}\right)^{1/2} \qquad (2)$$

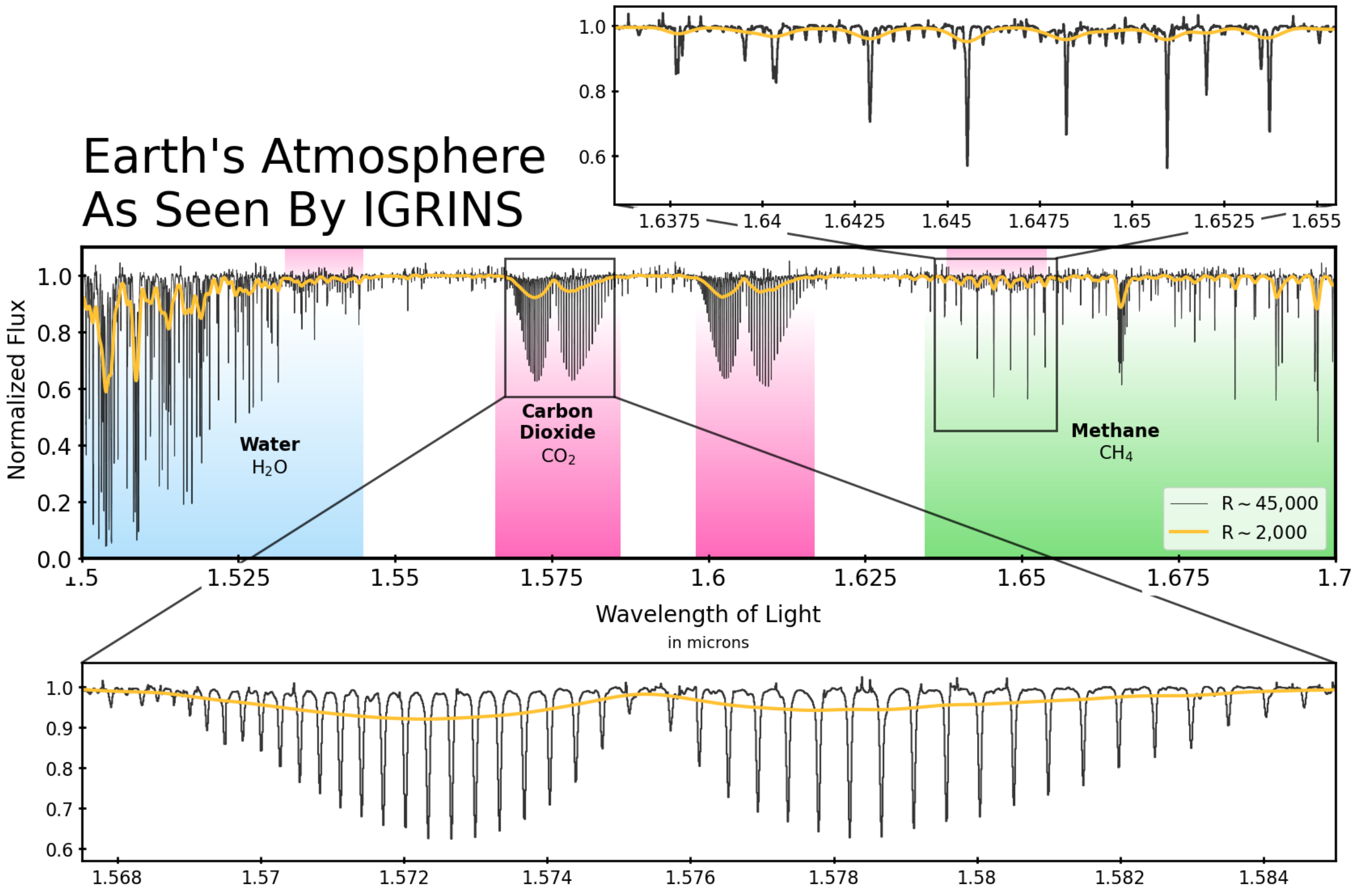

*Figure 1: Telluric spectrum observed at R = 45,000 with IGRINS (black) and the same spectrum smoothed to R = 2,000 (yellow). The middle panel labels the primary molecular constituents while the upper and lower panels show details of particular bands.*

Where $\eta_H$ and $\eta_L$ are the high- and low-resolution spectrograph throughputs and $R_H$ and $R_L$ are the resolving powers.

It is also interesting to consider regions with multiple lines from the same molecule. The feature in Figure 1 (bottom panel) stretches from 1.568 to 1.575 μm (about 4 resolution elements FWHM at R = 2,000) and has a depth ~ 5% of the continuum. The SNR for the feature detection at R = 2,000 is then $3.4 = \sqrt{4} \times 0.05 \times \frac{1125}{((1-0.05)\times 1125))^{\frac{1}{2}}}$. At R = 45,000, the $CO_2$ feature consists of ~12 unresolved features with an average depth of 0.3. Combining the detections of these features, the SNR for a detection is $8.8 = \sqrt{12} \times 0.3 \times \frac{100}{((1-0.3)\times 100)^{\frac{1}{2}}}$, 2.5 times greater than that of the low-resolution spectrum.

**Detection of weak features:** There are advantages of high-resolution beyond detectability. For many applications (for example, measuring isotopic ratios to determine the origin of molecular species), we will need to detect weak features in complex spectra. In the upper right panel of Figure 1 methane lines dominate the spectrum and, at R = 2,000, it is hard to detect any other species even at very high SNR. Spread throughout this region, however, are ~ 25 $CO_2$ features, each about 4% deep. Detection of 225 photons per R = 45,000 resolution element would permit a 3σ detection of $CO_2$.

**Demonstrated benefits:** Ground-based observations of hot Jupiters at high spectral resolution have demonstrated the advantage of placing incoming photons precisely in wavelength (e.g. Line et al. 2021, Brogi et al. 2023). More precise line wavelengths can eliminate confusion and remove ambiguities in identification of molecular species (e.g. Madhusudhan et al. 2025, Luque et al. 2025). Precise orbital velocities yield inclination-independent planetary masses (e.g. Brandt et al. 2019, Li et al. 2021). Phase offsets can reveal planetary winds (e.g. Shulyak et al. 2019) and line shapes can provide information about atmospheric structure (e.g. Chen et al. 2024).

These same advantages are present in observations of widely spaced massive planets like HR 8799 (Konopacky et al. 2013) and β Pic b (Snellen et al. 2014, Hoeijmakers et al. 2018) and in larger samples of such planets (Hoch et al. 2023, Xuan et al. 2024) when the resolving power is high enough to separate individual lines. In addition, using cross-correlation techniques on high spectral resolution data allows observers of these exoplanets to reach higher contrasts than are possible in the presence of speckle noise and other PSF irregularities (Ruffio et al. 2024, Xuan et al. 2024, Hoch et al. 2023).

**High-resolution helps in the real world:** The best coronographs will not completely eliminate spectral contamination by host stars. Since many of the most interesting planetary targets orbit Sun-like or cooler stars, the contaminating light will be spectrally complex. It will be considerably easier to remove the stellar spectrum or extract the planetary spectrum at high-resolution. This is particularly true for multi-epoch observations where planetary motion presents the target spectrum at independent positions with respect to the contaminating spectrum.

Even with a perfect coronograph, reflected starlight from planetary clouds will contaminate the intrinsic planetary spectrum. Higher-resolution helps in this case for single observations but even more for multi-epoch measurements since the reflected stellar spectrum Doppler shifts at twice the rate of the planetary absorption or emission spectrum.

## B. What about the background?

The line sensitivity advantage of high resolution persists at high background. Background photons affect the signal to noise in the same way that source photons do (Equation (1). As with the source photons, the number per resolution element scales inversely with R.

The present design of HWO affects the infrared sensitivity by limiting the longest wavelength where instruments can remain source-noise limited. If we make a fairly strict postulate, a 2000 second exposure and demand a signal to noise of 2 per resolution element against the continuum and that the telescope not decrease the SNR by more than 25%, we arrive at a requirement that the background be $\leq 0.0016/\eta_s$ (0.0032 for $\eta_s$=0.5) photons/sec/resolution element. Zodiacal backgrounds, both scattered and thermal, are too low to affect the infrared sensitivity at R>2000. Thermal emission from the telescope surfaces, however, imposes an upper limit on the operating wavelength of the spectrograph (Table 1).

| **Telescope Temperature (K)** | **Resolving Power $\lambda/\Delta\lambda$** | |
|---|---|---|
| | 2,000 | 45,000 |
| | $\lambda_{upper}$ ($\mu$m) | |
| **295** | 1.7 | 1.92 |
| **285** | 1.79 | 2.0 |
| **275** | 1.82 | 2.08 |

*Table 1: Limiting wavelengths for source-noise-limited operation at medium and high resolution.*

The values in Table 1 highlight the ability of a high-resolution spectrograph to remain in the source-noise limit to significantly longer wavelength at a given telescope operating temperature, a result that opens up more of the spectrum to efficient observation.

## C. Weighing the technical tradeoffs of high resolution

We present some of the potential technical objections to high-resolution infrared spectroscopy, explain how new technical solutions can mitigate negative effects, and assess the potential trade costs. Since the infrared cutoff wavelength for HWO instruments is still under consideration, we use a cutoff wavelength of 1.6 microns to provide some specificity.

**Throughput:** The band from 1.1 to 1.6 microns contains ~ 750 resolution elements at R = 2000 but ~ 17,000 resolution elements at R = 45,000. At reasonable sampling, the full spectrum fits on existing infrared focal plane arrays in a single order at R = 2,000. At R = 45,000, the spectrograph must operate in high order and employ cross-dispersion to place multiple orders onto the detector. The optical elements for cross-dispersion impose a throughput penalty. Apart from the additional grating, designs with comparable numbers of surfaces to existing designs with first-order reflective gratings are possible. The current design for the J, H- and K- band spectrographs in the GMTNIRS instrument (Figure 2) has only the immersion echelle and the cross disperser between the collimator and the camera. The measured throughput of our first-generation H- and K- band immersion echelles (Figure 3) is ~ 75% (Gully-Santiago et al. 2012). This efficiency is comparable to that of low-order surface-relief gratings that would serve as dispersers in low-resolution instruments and fine-tuned anti-reflection coatings could improve on the efficiency. Si grisms can serve as cross-dispersers in layouts like that of GMTNIRS (Fig. 2). Grisms similar to the parts UT produced for JWST NIRCam (Gully-Santiago et al. 2010) can have on-blaze throughputs >80%. When source photon noise predominates, the light from adjacent orders can be combined to achieve a flat throughput close to the blaze maximum. It is likely that a properly designed cross-dispersed instrument can have a throughput >75% of the throughput of a low-resolution instrument. For R = 45,000 vs. R = 2,000, therefore, the SNR advantage would still be $\geq 3.5\times$ for single unresolved lines.

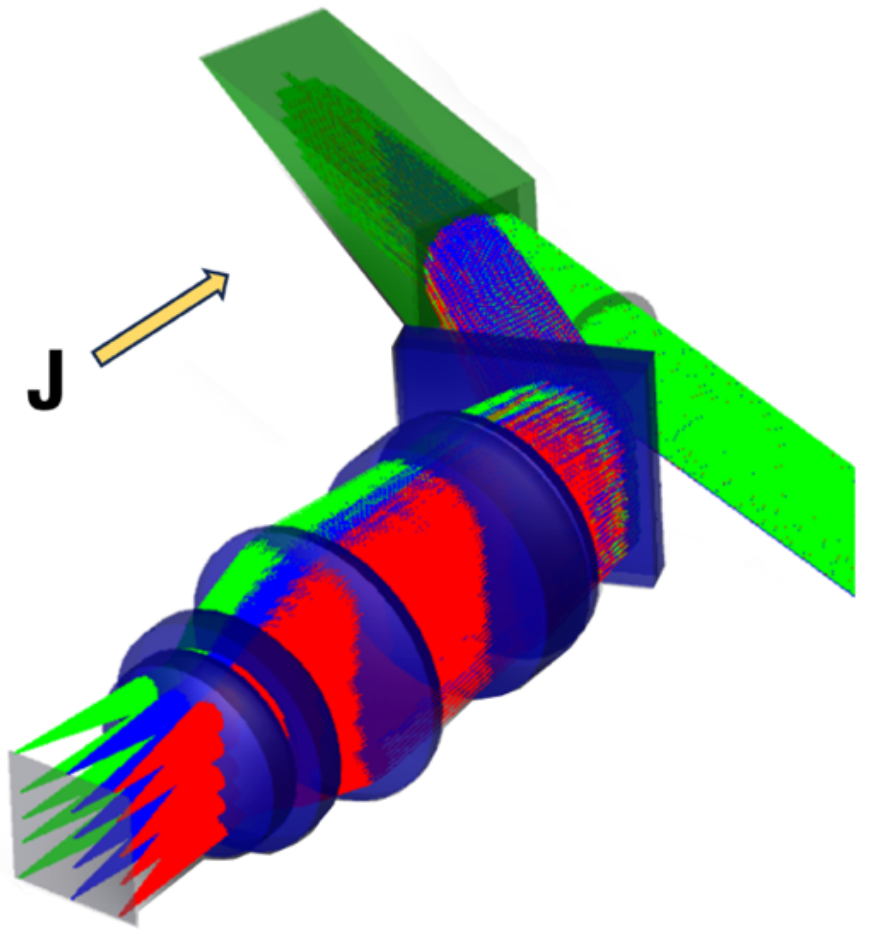

*Figure 2: J-band cross-dispersed unit for the GMTNIRS spectrograph. An R3 silicon immersion grating (top) serves as the high-resolution disperser and a VPH transmission grating (blue square at center) as the cross-disperser. This design can produce spectra of the entire J band at R = 75,000. The collimated beam size is 25 mm and the entire unit is less than 175 mm x 175 mm.*

**Detector noise and dark current:** Spectrograph designers typically sample the slit image in the focal plane with m = 2-4 pixels, independent of resolving power. The photon-equivalent detector noise and dark current are therefore the same at all resolutions while the depth of the line increases linearly with resolving power. When the noise in the system consists of a combination of source, background, dark current and detector noise, the SNR advantage factor at high-resolution is lower than in the source-noise-only limit and the advantage disappears as the read noise and/or dark current begin to dominate. The left-hand term in the denominator of Eq. 1 is source and background photon noise while the right-hand term is detector dark current and read noise. For an instrument to remain in the source plus background noise limited regime, the number of detected photons per pixel needs to be about $4 \times$ the square of the read noise and $4 \times$ the number of dark current electrons. When the read noise is low, even detections at low SNR can occur in the source noise limit. Typical read noises for the integrating source follower per detector HgCdTe arrays in use on today's instruments are around 5 electrons for long exposures and multiple reads, meaning there would need to be ~ 100 photoelectrons per pixel (~ 300 per resolution element) to reach the source noise limit. Recent developments in avalanche photodiode arrays have demonstrated effective read noise less than one electron and dark currents below $10^{-3}$ e/s/pixel with comparable quantum efficiencies (Finger et al. 2016, Feautrier &

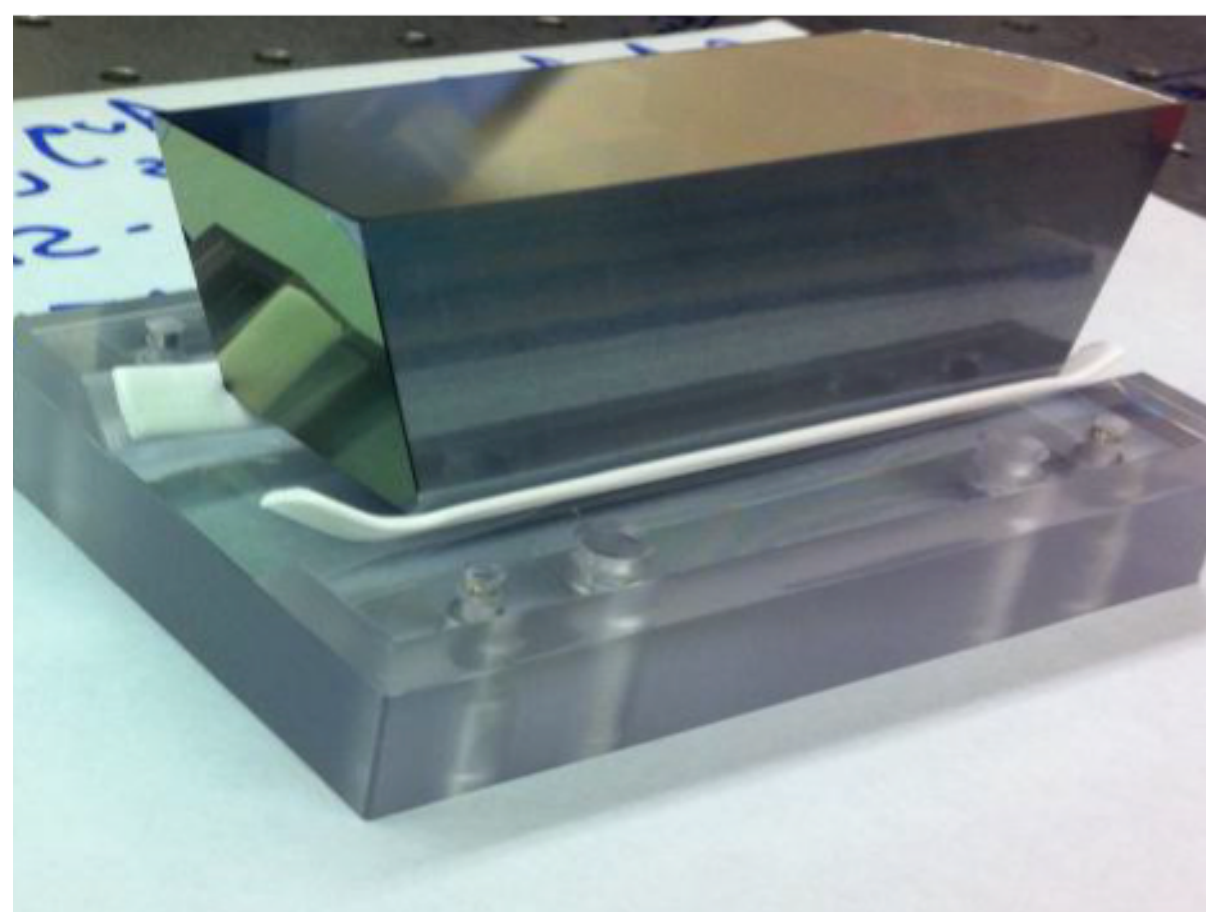

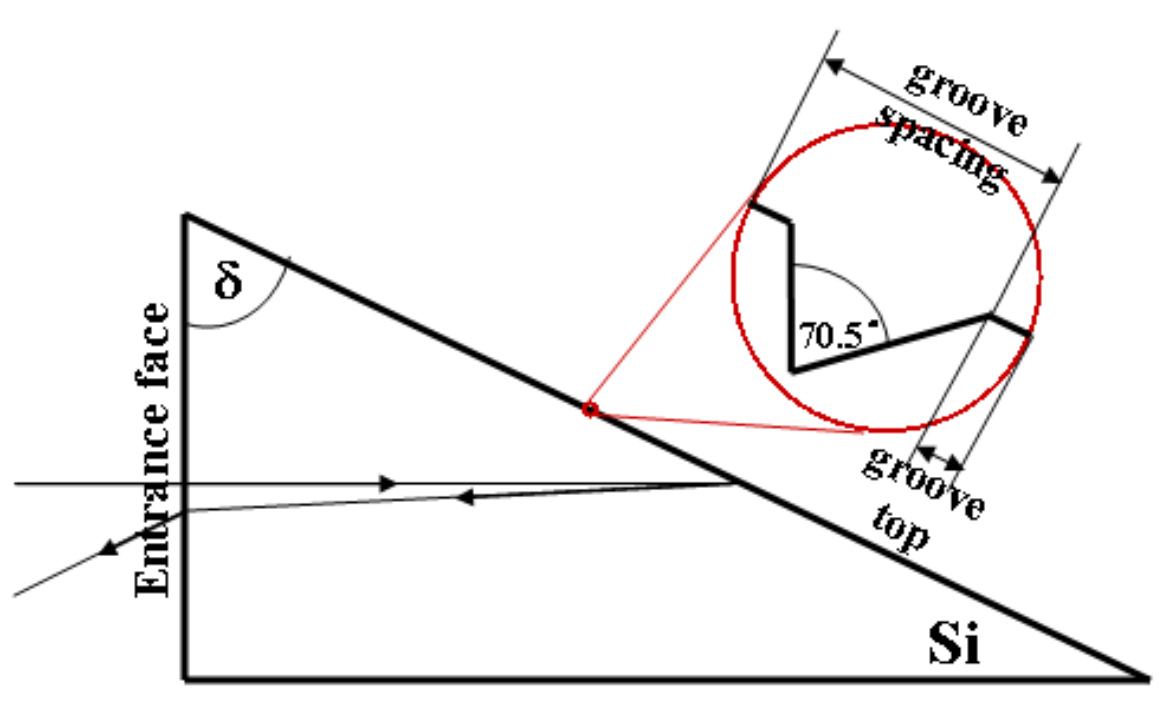

*Figure 3: (Left) Immersion echelle grating for the R = 45,000 IGRINS spectrometer. The AR coated entrance / exit face (27 mm x 27 mm) is on the left and the Al coated grating surface is on the top. (Right) Diagram illustrating the light path in an immersion grating.*

Gach 2022, Claveau et al. 2024). With this performance, source photon noise will dominate spectroscopic observations with only ~ 4 detected photons per pixel and the comparisons in the "Raw detectability" section will be valid for all (>3$\sigma$) detections.

**Detector real estate:** With a custom cross-disperser, it is possible to place the whole spectrum of interest on a 2k x 2k array with an appropriate slit length. We can apportion the 17,000 resolution elements required to observe the 1.0-1.6 $\mu$m spectral region at R = 45,000 across ~ 25 orders at 2.5 pixels / resolution element. The slit length could then be ~ 35 pixels, or ~ 14 times the slit width. These results imply that it is easily possible to build a high-resolution instrument for the proposed near-IR band that would cover the entire band in a single exposure and require no moving parts.

**Size and weight:** The spectrograph unit shown in Figure 2, which is <200mm x 200mm, is sized for a slit that is ~ 4$\lambda$/D. The overall size of the instrument scales roughly linearly with the slit size. An appropriate near-IR high-resolution unit for HWO could be built using Si diffractive optics at a minimal cost in size and weight.

**Do we need two spectrographs?** Some sources only have broad lines. It is worth examining whether the mission needs a second, low-resolution instrument to observe these objects. In the source-noise limit, one can sum up a series of high-resolution channels to produce a low-resolution spectrum with less noise per resolution element. The only loss comes from the somewhat lower throughput. The other issue is whether we can stitch together the more complex cross-dispersed spectra without introducing systematics. The SS 433 result of Robinson et al. 2017 (Figure 4) is a feasibility demonstration for the case of IGRINS, a large-grasp spectrograph with no moving parts.

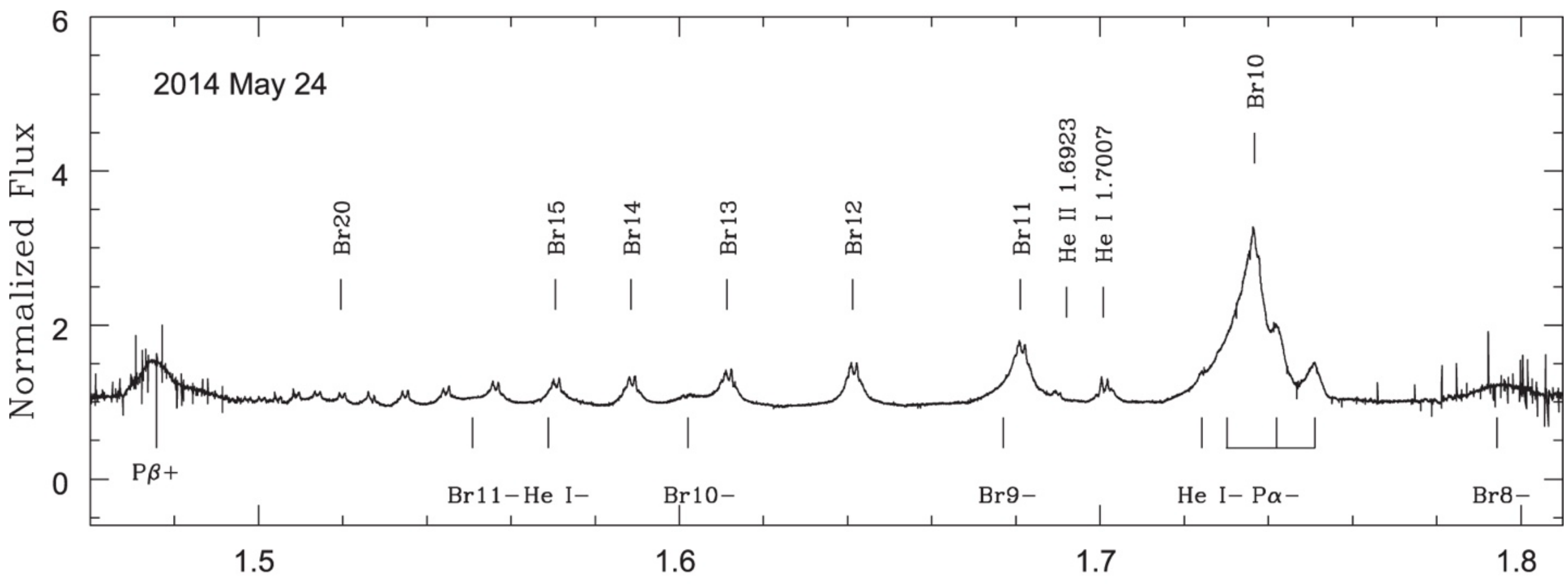

*Figure 4: IGRINS R = 45,000 H-band spectrum of the X-ray binary SS 433 (Robinson et al. 2017). The spectrum demonstrates how >20 spectral orders can be seamlessly stitched together to produce a spectrum that includes broad features.*

## D. What needs to be done?

Many of the elements for a high-resolution instrument spanning 1-1.6 μm are already mature, but critical work remains. Silicon grisms constructed by the University of Texas group have flown as dispersive elements in the JWST NIRCam spectrograph (Greene et al. 2016). Three ground-based instruments, IGRINS (Park et al. 2014), iSHELL (Rayner et al. 2022), and IGRINS-2 (Choi et al. 2025), which have produced hundreds of published papers, use silicon immersion gratings from UT as their primary dispersers.

The size, incidence angle and groove pitch of the immersion echelles and the cross dispersers all affect their manufacturability and the effective maturity of the instrument. It is therefore important to flow down an initial set of science requirements into a set of proto-specifications that we can use as the basis for an initial optical design. With that design in hand, we can produce a set of test immersion gratings to evaluate performance to understand limitations and can ultimately use in brass boards and more mature prototypes. Grisms are significantly easier to manufacture than immersion gratings but also need further development. We would need to carry out a rigorous coupled wave analysis of the design to arrive at a structure that maximizes throughput across the desired band. We would additionally need to experiment with antireflection coatings for the grooved surface to understand how to optimize grating/grism efficiency.

Avalanche photodiode arrays are rapidly improving in dark current and format size and offer promise of meeting the requirements for a high-resolution instrument. It will take some careful interaction with vendors to test and optimize the devices for the low light levels present in a high dispersion spectrograph; in particular, the optimization of dark current while simultaneously maintaining low read noise. We would need also to test the devices in a realistic spectrograph to understand how pixel crosstalk and persistence affect the fidelity of astronomical spectrographs.

Ultimately, we would need to build a full-up prototype of the instrument. Given its likely size and weight, the possibility exists of flying an early version of this prototype on a smaller satellite to bring the entire instrument to high TRL pre-HWO launch.

**References**

Brandt, T., Dupuy, T., & Bowler, B. (2019). Precise Dynamical Masses of Directly Imaged Companions from Relative Astrometry, Radial Velocities, and Hipparcos-Gaia DR2 Accelerations. *The Astronomical Journal, 158(4), 140*. https://doi.org/10.3847/1538-3881/ab04a8

Brogi, M., Emeka-Okafor, V., Line, M., Gandhi, S., Pino, L., Kempton, E.R., Rauscher, E., Parmentier, V., Bean, J., Mace, G., Cowan, N., Shkolnik, E., Wardenier, J., Mansfield, M., Welbanks, L., Smith, P., Fortney, J., Birkby, J., Zalesky, J., Dang, L., Patience, J., & Desert, J.M. (2023). The Roasting Marshmallows Program with IGRINS on Gemini South I: Composition and Climate of the Ultrahot Jupiter WASP-18 b. *The Astronomical Journal, 165(3), 91*. https://doi.org/10.3847/1538-3881/acaf5c

Chen, X., Biller, B., Vos, J., Crossfield, I., Mace, G., Hood, C., Tan, X., Allers, K., Martin, E., Bubb, E., Fortney, J., Morley, C., & Hammond, M. (2024). Global weather map reveals persistent top-of-atmosphere features on the nearest brown dwarfs. *Monthly Notices of the Royal Astronomical Society, 533(3), 3114-3143*. https://doi.org/10.1093/mnras/stae1995

Choi, Y.H., Jeong, U., Lee, J.J., Kim, H.J., Oh, H., Park, C., Kye, C., Finnerty, L., Line, M., Kanumalla, K., Sanchez, J., Smith, P., Kim, S., Lee, H.I., Park, W., Yu, Y., Kim, Y., Chun, M.Y., Oh, J., Lee, S., Jang, J.G., Jang, B.H., Seong, H., Brooks, C., Mace, G., Lee, H., Good, J., Jaffe, D., Kim, K.M., Yuk, I.S., Hwang, N., Park, B.G., Kim, H., Chinn, B., Ramos, F., Prado, P., Diaz, R., White, J., Tapia, E., Olivares, A., Oyarzun, V., Kurz, E., Stecher, H., Quiroz, C., Arriagada, I., Hayward, T., Suh, H., Miller, J., Xu, S., Farina, E., Figura, C., Mocnik, T., Hartman, Z., Rawlings, M., Stephens, A., Miller, B., Labrie, K., Hirst, P., & Lee, B.C. (2025). An Early Look at the Performance of IGRINS-2 at Gemini-North with Application to the Ultrahot Jupiter, WASP-33 b. *The Astronomical Journal, 170(4), 238*. https://doi.org/10.3847/1538-3881/aded8c

Claveau, C.A., Bottom, M., Jacobson, S., Hodapp, K., Huber, G., Newland, M., Walk, A., Loose, M., Baker, I., Zemaityte, E., Hicks, M., Barnes, K., Powell, R., Bradley, R., & Moore, E. (2024). Progress towards a megapixel linear-mode avalanche photodiode array for ultra-low background shortwave infrared astronomy. *arXiv e-prints*, arXiv:2411.09185. https://doi.org/10.48550/arXiv.2411.09185

Cubillos, P., Brogi, M., Muñoz, A., Fossati, L., Blecic, J., Saikia, S., Bourrier, V., Caballero, J., Cabrera, Chiavassa, A., Fludracc, A., Gkouveliscc, L., Grenfellcc, J., Guedelcc, M., Hazracc, G., Labianoc, A., Lendl, M., Rodgers-Lee, D., Salvador, Schroetter, I., Strugarek, A., Taysum, B., Vidotto, A., & Wilson, T. (2026). High-resolution Ultraviolet-to-Near-Infrared Characterization of Exoplanet Atmospheres with HWO. In *Astronomical Society of the Pacific Conference Series* (pp. 545). https://doi.org/10.48550/arXiv.2507.03060

Feautrier, P., & Gach, J.L. (2022). Last performances improvement of the C-RED One camera using the 320x256 e-APD infrared Saphira detector. In *Optical and Infrared Interferometry and Imaging VIII* (pp. 121832E). Finger, G., Baker, I., Alvarez, D. et al. 2016, Proc. SPIE 9909, 990912. https://doi.org/10.1117/12.2630860

Greene, T., Chu, L., Egami, E., Hodapp, K., Kelly, D., Leisenring, J., Rieke, M., Robberto, M., Schlawin, E., & Stansberry, J. (2016). Slitless spectroscopy with the James Webb Space Telescope Near-Infrared Camera (JWST NIRCam). In *Space Telescopes and Instrumentation 2016: Optical, Infrared, and Millimeter Wave* (pp. 99040E). https://doi.org/10.1117/12.2231347

Gully-Santiago, M., Wang, W., Deen, C., Kelly, D., Greene, T., Bacon, J., & Jaffe, D. (2010). High-performance silicon grisms for 1.2-8.0 μm: detailed results from the JWST-NIRCam devices. In *Modern Technologies in Space- and Ground-based Telescopes and Instrumentation* (pp. 77393S). https://doi.org/10.1117/12.857568

Gully-Santiago, M., Wang, W., Deen, C., & Jaffe, D. (2012). Near-infrared metrology of high-performance silicon immersion gratings. In *Modern Technologies in Space- and Ground-based Telescopes and Instrumentation II* (pp. 84502S). https://doi.org/10.1117/12.926434

Hoch, K., Konopacky, Q., Theissen, C., Ruffio, J.B., Barman, T., Rickman, E., Perrin, M., Macintosh, B., & Marois, C. (2023). Assessing the C/O Ratio Formation Diagnostic: A Potential Trend with Companion Mass. *The Astronomical Journal, 166(3), 85.* https://doi.org/10.3847/1538-3881/ace442

Hoeijmakers, H., Schwarz, H., Snellen, I., de Kok, R., Bonnefoy, M., Chauvin, G., Lagrange, A., & Girard, J. (2018). Medium-resolution integral-field spectroscopy for high-contrast exoplanet imaging. Molecule maps of the β Pictoris system with SINFONI. *Astronomy & Astrophysics, 617, A144.* https://doi.org/10.1051/0004-6361/201832902

Jakobsen, P., Ferruit, P., Oliveira, C., Arribas, S., Bagnasco, G., Barho, R., Beck, T., Birkmann, S., Boker, T., Bunker, A., Charlot, S., de Jong, P., de Marchi, G., Ehrenwinkler, R., Falcolini, M., Fels, R., Franx, M., Franz, D., Funke, M., Giardino, G., Gnata, X., Holota, W., Honnen, K., Jensen, P., Jentsch, M., Johnson, T., Jollet, D., Karl, H., Kling, G., Köhler,

J., Kolm, M.G., Kumari, N., Lander, M., Lemke, R., Lopez-Caniego, M., Lützgendorf, N., Maiolino, R., Manjavacas, E., Marston, A., Maschmann, M., Maurer, R., Messerschmidt, B., Moseley, S., Mosner, P., Mott, D., Muzerolle, J., Pirzkal, N., Pittet, J.F., Plitzke, A., Posselt, W., Rapp, B., Rauscher, B., Rawle, T., Rix, H.W., Rödel, A., Rumler, P., Sabbi, E., Salvignol, J.C., Schmid, T., Sirianni, M., Smith, C., Strada, P., te Plate, M., Valenti, J., Wettemann, T., Wiehe, T., Wiesmayer, M., Willott, C., Wright, R., Zeidler, P., & Zincke, C. (2022). The Near-Infrared Spectrograph (NIRSpec) on the James Webb Space Telescope. I. Overview of the instrument and its capabilities. *Astronomy & Astrophysics, 661, A80*. https://doi.org/10.1051/0004-6361/202142663

Konopacky, Q., Barman, T., Macintosh, B., & Marois, C. (2013). Detection of Carbon Monoxide and Water Absorption Lines in an Exoplanet Atmosphere. *Science, 339(6126), 1398-1401*. https://doi.org/10.1126/science.1232003

Li, Y., Brandt, T., Brandt, G., Dupuy, T., Michalik, D., Jensen-Clem, R., Zeng, Y., Faherty, J., & Mitra, E. (2021). Precise Masses and Orbits for Nine Radial-velocity Exoplanets. *The Astronomical Journal, 162(6), 266*. https://doi.org/10.3847/1538-3881/ac27ab

Line, M., Brogi, M., Bean, J., Gandhi, S., Zalesky, J., Parmentier, V., Smith, P., Mace, G., Mansfield, M., Kempton, E.R., Fortney, J., Shkolnik, E., Patience, J., Rauscher, E., Desert, J.M., & Wardenier, J. (2021). A solar C/O and sub-solar metallicity in a hot Jupiter atmosphere. *Nature, 598(7882), 580-584*. https://doi.org/10.1038/s41586-021-03912-6

Luque, R., Piaulet-Ghorayeb, C., Radica, M., Xue, Q., Zhang, M., Bean, J., Samra, D., & Steinrueck, M. (2025). Insufficient evidence for DMS and DMDS in the atmosphere of K2-18 b: From a joint analysis of JWST NIRISS, NIRSpec, and MIRI observations. *Astronomy & Astrophysics, 700, A284*. https://doi.org/10.1051/0004-6361/202555580

Madhusudhan, N., Constantinou, S., Holmberg, M., Sarkar, S., Piette, A., & Moses, J. (2025). New Constraints on DMS and DMDS in the Atmosphere of K2-18 b from JWST MIRI. *The Astrophysical Journal Letters, 983(2), L40*. https://doi.org/10.3847/2041-8213/adc1c8

Park, C., Jaffe, D., Yuk, I.S., Chun, M.Y., Pak, S., Kim, K.M., Pavel, M., Lee, H., Oh, H., Jeong, U., Sim, C., Lee, H.I., Nguyen Le, H., Strubhar, J., Gully-Santiago, M., Oh, J., Cha, S.M., Moon, B., Park, K., Brooks, C., Ko, K., Han, J.Y., Nah, J., Hill, P., Lee, S., Barnes, S., Yu, Y., Kaplan, K., Mace, G., Kim, H., Lee, J.J., Hwang, N., & Park, B.G. (2014). Design and early performance of IGRINS (Immersion Grating Infrared Spectrometer). In *Ground-based and Airborne Instrumentation for Astronomy V* (pp. 91471D. https://doi.org/10.1117/12.2056431

Rayner, J., Tokunaga, A., Jaffe, D., Bond, T., Bonnet, M., Ching, G., Connelley, M., Cushing, M., Kokubun, D., Lockhart, C., Vacca, W., & Warmbier, E. (2022). iSHELL: a 1-5 micron R = 80,000 Immersion Grating Spectrograph for the NASA Infrared Telescope Facility.

*Publications of the Astronomical Society of the Pacific, 134(1031), 015002*. https://doi.org/10.1088/1538-3873/ac3cb4

Robinson, E., Froning, C., Jaffe, D., Kaplan, K., Kim, H., Mace, G., Sokal, K., & Lee, J.J. (2017). The Spectrum of SS 433 in the H and K Bands. *The Astrophysical Journal, 841(2), 79*. https://doi.org/10.3847/1538-4357/aa6f0c

Ruffio, J.B., Perrin, M., Hoch, K., Kammerer, J., Konopacky, Q., Pueyo, L., Madurowicz, A., Rickman, E., Theissen, C., Agrawal, S., Greenbaum, A., Miles, B., Barman, T., Balmer, W., Llop-Sayson, J., Girard, J., Rebollido, I., Soummer, R., Allen, N., Anderson, J., Beichman, C., Bellini, A., Bryden, G., Espinoza, N., Glidden, A., Huang, J., Lewis, N., Libralato, M., Louie, D., Sohn, S., Seager, S., Marel, R., Wakeford, H., Watkins, L., Ygouf, M., & Mountain, C. (2024). JWST-TST High Contrast: Achieving Direct Spectroscopy of Faint Substellar Companions Next to Bright Stars with the NIRSpec Integral Field Unit. *The Astronomical Journal, 168(2), 73*. https://doi.org/10.3847/1538-3881/ad5281

Shulyak, D., Rengel, M., Reiners, A., Seemann, U., & Yan, F. (2019). Remote sensing of exoplanetary atmospheres with ground-based high-resolution near-infrared spectroscopy. *Astronomy and Astrophysics, 629, A109*. https://doi.org/10.1051/0004-6361/201935691

Snellen, I. (2025). Exoplanet Atmospheres at High Spectral Resolution. *Annual Review of Astronomy and Astrophysics, 63(1), 83-125*. https://doi.org/10.1146/annurev-astro-052622-031342

Snellen, I., Brandl, B., de Kok, R., Brogi, M., Birkby, J., & Schwarz, H. (2014). Fast spin of the young extrasolar planet β Pictoris b. *Nature, 509(7498), 63-65*. https://doi.org/10.1038/nature13253

Strugarek, A., Berdyugina, S., Bourrier, V., Caballero, J., Chebly, J., Fares, R., Fludra, A., Fossati, L., Muñoz, A., Gkouvelis, L., Gourves, C., Grenfell, J., Kavanagh, R., Kislyakova, K., Lamy, L., Lanza, A., Moutou, C., Nandy, D., Neiner, C., Oklopčić, A., Paul, A., Réville, V., Rodgers-Lee, D., Shkolnik, E., Turner, J., Vidotto, A., Yang, F., & Zarka, P. (2026). Detecting and Characterising the Magnetic Field of Exoplanets. In *Astronomical Society of the Pacific Conference Series* (pp. 565). https://doi.org/10.48550/arXiv.2507.02010

Wakeford, H., Mayorga, L., Barstow, J., Batalha, N., Carone, L., Casewell, S., Karalidi, T., Kataria, T., May, E., & Min, M. (2025). Characterizing the Dynamics and Chemistry of Transiting Exoplanets with the Habitable World Observatory. *arXiv e-prints*, arXiv:2506.22839. https://doi.org/10.48550/arXiv.2506.22839

Xu, S., Barstow, M., Buchan, A., Le Bourdais, É., & Dufour, P. (2025). White dwarfs as probes of extrasolar planet compositions and fundamental astrophysics. *arXiv e-prints*, arXiv:2507.03029. https://doi.org/10.48550/arXiv.2507.03029

Xuan, J., Hsu, C.C., Finnerty, L., Wang, J., Ruffio, J.B., Zhang, Y., Knutson, H., Mawet, D., Mamajek, E., Inglis, J., Wallack, N., Bryan, M., Blake, G., Molliere, P., Hejazi, N., Baker,

A., Bartos, R., Calvin, B., Cetre, S., Delorme, J.R., Doppmann, G., Echeverri, D., Fitzgerald, M., Jovanovic, N., Liberman, J., Lopez, R., Morris, E., Pezzato, J., Sappey, B., Schofield, T., Skemer, A., Wallace, J., Wang, J., Agrawal, S., & Horstman, K. (2024). Are These Planets or Brown Dwarfs? Broadly Solar Compositions from High-resolution Atmospheric Retrievals of \ensuremath~10–30 $M_{Jup}$ Companions. *The Astrophysical Journal, 970(1), 71*. https://doi.org/10.3847/1538-4357/ad4796